\documentstyle[12pt]{article}
\makeatletter
\newcommand{\singlespacing}{\let\CS=\@currsize\renewcommand{\baselinestretch}
{1.0}\tiny\CS}
\newcommand{\doublespacing}{\let\CS=\@currsize\renewcommand{\baselinestretch}
{1.5}\tiny\CS}

\begin{document}

{\noindent {\huge {\bf Realization of Optimal Disentanglement 
by Teleportation via Separable}}} 
\begin{center}
{\huge {\bf Channel}}
\end{center}

\begin{center}
\vspace{0.5cm}
{\bf Sibasish Ghosh}$^a$\footnote{E--mail address : res9603@www.isical.ac.in},
{\bf Guruprasad Kar}$^a$\footnote{E--mail address : gkar@www.isical.ac.in}, {\bf Anirban Roy}$^a$\footnote{E--mail address :
res9708@www.isical.ac.in}, {\bf Debasis Sarkar}$^b$ and {\bf Ujjwal
Sen}$^c$\footnote{E--mail address : dhom@bosemain.boseinst.ernet.in}
\end{center}

\vspace{0.4cm}
{\noindent 
$^a${\it Physics and Applied Mathematics Unit, Indian Statistical Institute, 
203 B.T. Road, Calcutta -- 700 035, India}.\\
$^b${\it Department of Applied Mathematics, University of Calcutta, 92 A.P.C.
Road, Calcutta -- 700 009, India}.\\
$^c${\it Department of Physics, Bose Institute, 93/1 A.P.C. Road, Calcutta -- 700 009, India}.}

\begin{center}
\vspace{1cm}
{\bf Abstract}
\end{center}

\singlespacing
{\small {We discuss here the best disentanglement processes of states of two
two--level systems which belong to
(i) the universal set, (ii) the set in which the states of one party lie on a
single great circle of the Bloch sphere, and (iii) the set in which the states
of one party commute with each other, by teleporting the states of one party
(on which the disentangling machine is acting) through three particular type 
of separable channels, each of which is a mixture of Bell states. In the
general scenario, by teleporting one party's state of an arbitrary entangled
state of two two--level parties through some mixture of Bell states, we have
shown that this entangled state can be made separable by using a
physically realizable map $\tilde{V}$, acting on one party's states, if
$\tilde{V} (I) = I, \tilde{V} ({\sigma}_j) = {\lambda}_j {\sigma}_j$, where
${\lambda}_j \ge 0$ (for $j = 1, 2, 3$), and ${\lambda}_1 + {\lambda}_2 +
{\lambda}_3 \le 1$.}}

\doublespacing
\section{Introduction}
Entanglement plays one of the most fundamental roles in quantum theory. Particularly, in quantum 
information theory, it plays such roles which are highly counter--intuitive in the classical 
world. In quantum theory, one is allowed to do only measurement, unitary
operations on a system, and 
classical communications (e.g., by using phone calls) between two or more
separated parties. And if two or more parties share 
some entanglement, the outcomes of the measurements on the state of the particle possessed by one 
party can not be specified independently of the parameters of the measurements on the states of 
the particles possessed by the other parties. A fundamental principle in quantum theory is that one can not create entanglement by using local operations (e.g., local 
measurements, local unitary operations) and classical communications (LOCC). So there remains the possibility of disentanglement of an entangled state into a separable state when one 
uses LOCC. And like other no--go theorems (e.g., no--cloning 
theorem \cite{WZ}, no--deletion theorem \cite{PB}, no--broadcasting theorem
\cite{B}), it was recently shown that the no--disentanglement theorem
\cite{Terno, Mor} also holds in quantum information theory. Thus one can not disentangle an 
arbitrarily given state of two two--level systems into a separable state, keeping the local density matrices 
intact. This type of disentanglement is called {\it exact}, as one has to keep
the local density matrices intact. So, just as in the case of optimal universal cloning \cite{BH, Bruss}, one may ask 
how far can we go by disentangling {\it inexactly}, {\it i.e.}, how close can the output reduced 
density matrices be taken to the corresponding input reduced density matrices. Both symmetric 
and asymmetric optimal universal disentangling machines have been obtained, by using local 
operations \cite{Bandyo, Ghosh, Zhang}. If ${\eta}_1$ and ${\eta}_2$ are the reduction 
factors (reducing the Bloch vectors of the input reduced density matrices corresponding to the 
party 1 and party 2 by ${\eta}_1$ and ${\eta}_2$ respectively), it has been shown that 
${\eta}_1 {\eta}_2 \le 1/3$. In the symmetric case, the result $\eta \le 1/{\sqrt{3}}$ has 
been obtained in ref. \cite{Bandyo}, by using approximate cloning machines on
the two parties separately. What is the physical significance of this value $1/{\sqrt{3}}$, in the 
case of optimal inexact disentanglement? Here one remembers the result that the $1 \rightarrow 2$ 
optimal universal cloning machines can be achieved by using the principle of no--superluminal 
signalling \cite{Gisin, Ghosh1, B1}. 
In this paper, we provide a possible explanation of this value $1/3$ (in the
case of the most 
asymmetric disentanglement, {\it viz}., ${\eta}_1 = 1, {\eta}_2 = 1/3$), by
teleporting one party of an arbitrarily given state of two two--level systems through a Werner channel \cite{Werner}.

Is there any set of (two--qubit) states, which can be disentangled exactly? Mor and Terno 
\cite{MT} provided a sufficent condition in this case : if the reduced density matrices of one of the two parties of 
each of the bipartite states from the set, commute with each other, then this
set can be exactly disentangled. Taking advantage of the fact that any known
set of commuting (two--level) states can be teleported through a 
separable channel, we realize the above phenomenon of exact disentanglement by
teleporting the state of one party through a separable channel \cite{Ghosh3}.

Between the whole Bloch sphere and any diameter of it (along which the Bloch vectors of any given 
set of commuting states lie), there are infinitely many subsets. But any great circle of the Bloch 
sphere is of particular interest : (i) any great circle is the largest set of
Bloch vectors that can be exactly flipped \cite{Pati, GisinPopescu}, and this fact can be used for remote (exact) state preparation 
\cite{Bennett}, (ii) a product state of two parallel qubits and the 
corresponding product state of two antiparallel qubits contain equivalent information about 
the qubit, if the Bloch vectors of all these qubits lie on the same great circle \cite{Ghosh2},
(iii) the optimal universal $1 \rightarrow 2$ (isotropic) cloning machine of the equator (on which
the states $|0\rangle$, $|1\rangle$, $(1/{\sqrt{2}}) (|0\rangle \pm |1\rangle)$ lie) is 
the same as that of the set $\{|0\rangle, |1\rangle, (1/{\sqrt{2}}) (|0\rangle \pm |1\rangle) \}$
\cite{B3}. We consider here the set of all entangled states, where all the Bloch vectors of the 
reduced density matrices of one side lie on the disc of a great circle. By
applying a map (on the side whose density matrices lie on the said disc), 
which preserves the isotropy of the Bloch vectors on this ``great disc" \cite{greatdisc}, we obtained the 
optimal disentangling machine corresponding to this set. Again here, we provide
a physical significance of 
the existence of this disentangling machine, by teleporting the states of that particular side 
of the entangled states from the above set, through a separable channel (a
specific mixture of the Bell states).

Here we also deal all of the above three cases of disentanglement (together with
their generalization) in a common footing : by teleporting states of a system 
through any mixture of the four Bell states (used as the channel). And we 
discuss about the property of the isotropy preserving map, realizing this 
teleportation process.

The paper is arranged as follows. In section 2, we first define the
disentanglement process. Then we describe briefly the optimal universal 
disentanglement process and the corresponding isotropy preserving maps. In 
section 3, we describe the sufficient criterion for exact 
disentanglement -- in a very simple manner. In section 4, we describe the optimal (isotropic) 
disentanglement process of a set of states, where the Bloch vector of the reduced density matrix 
of one side of each of these states lie on the same great disc. In section 5, we consider the 
teleportation process of one party's state, taken at random from a given set of two--qubit 
entangled states, through a channel, which is a mixture of the four Bell states. Using the 
separability of the channel, we find the property of the linear map 
and associate it with disentanglement of the set. And then, 
the particular types of disentanglement, described in the earlier sections, are shown as special 
cases of this disentanglement scenario. In section 6, we draw the conclusions. 

Before the starting of the next section, let us clarify some necessary ideas
which will be required in this paper. Let us take ${\cal D}$ to be the set of
all density matrices $(1/2) (I + \vec{r} . \vec{\sigma})$ (where $\vec{r}$ is a
real three--dimensional vector with $|\vec{r}| \le 1$, and ${\sigma}_x$,
${\sigma}_y$, ${\sigma}_z$ are the standard three Pauli spin matrices) of a 
two--level system. Here the support of each of these density matrices is the
two--dimensional Hilbert space ${C\!\!\!\!I}^2$. Let us take ${\cal L}$ to be the
set of all linear operators having both domain and range as ${C\!\!\!\!I}^2$. Thus
${\cal D}$ is a subset of ${\cal L}$. We take ${\cal D}_1$ to be any non--empty
subset of ${\cal D}$. Also let $\tilde{V} : {\cal D}_1 \rightarrow {\cal L}$ be
any linear map.

{\noindent {\bf Definition 1 :} We call the map $\tilde{V}$ to be {\it
physical} if (i) ${\cal D}_1 = {\cal D}$, and (ii) the range $\tilde{V} ({\cal
D}_1)$ of the map $\tilde{V}$ is a subset of ${\cal D}$.}

{\noindent {\bf Definition 2 :} We call the map $\tilde{V}$ to be {\it
realizable} if there exists an ancilla system ${\cal A}$ described by the Hilbert space
${\cal H}_{\cal A}$, and an unitary operator $U : {C\!\!\!\!I}^2 \otimes {\cal
H}_{\cal A} \rightarrow {C\!\!\!\!I}^2 \otimes {\cal H}_{\cal A}$ such that ${\rm
Tr}_{\cal A} [U {\rho} \otimes P[|M\rangle] U^{\dag}] = \tilde{V} (\rho)$ for
every $\rho \in {\cal D}_1$, where $|M\rangle$ is a fixed state in ${\cal
H}_{\cal A}$.}

One should note that every physical map is a realizable map also.

\section{Optimal universal disentanglement process}
Consider a (non--empty) set ${\cal S}$ of some states of two two--level systems
shared between $A$ and $B$. If, by any process (allowed by quantum mechanics), every state
$\rho$ from the set ${\cal S}$ becomes a separable state ${\rho}^{\prime}$ 
(which is being possessed now by $C$ and 
$D$, say, where $(A, B)$ may or may not be equal to $(C, D)$), where the 
reduced density matrices of both the particles remain unchanged in this 
process (except the possible changes $A \rightarrow C$ and $B \rightarrow D$ 
in location), this process is then called {\it exact} 
disentanglement of the set ${\cal S}$.    
  
A fundamental property of quantum mechanics is that exact disentanglement of 
the set ${\cal S}_{universal}$ of all states of two two--level systems is 
impossible \cite{Terno, Mor}. But how well can we retain the local density 
matrices ({\it i.e.}, reduced density matrices) of a bipartite state $\rho$, 
selected at random from the set ${\cal S}_{universal}$, by {\it disentangling} 
this state? More specifically, if, before disentanglement of the bipartite 
state $\rho$, ${\rho}_1 = (1/2) (I + {\bf n}.{\bf {\sigma}})$ 
(${\rho}_2 = (1/2) (I + {\bf m}.{\bf {\sigma}})$) is the reduced density matrix 
of the first (second) party, where ${\bf n}$ (${\bf m}$) is the Bloch vector 
corresponding to ${\rho}_1$ (${\rho}_2$) with $|{\bf n}| \le 1$ 
($|{\bf m}| \le 1$), and if, after disentanglement of the state $\rho$ into 
some bipartite separable state ${\rho}^{\prime}$, 
${\rho}^{\prime}_1 = (1/2) (I + {\eta}_1 {\bf n}.{\bf {\sigma}})$ (${\rho}^{\prime}_2 = (1/2) (I + {\eta}_2 {\bf m}.{\bf {\sigma}})$) 
is the reduced density matrix of the first (second) party (where the 
identities of the particles may not be retained),  where 
$0 \le {\eta}_1 \le 1$ ($0 \le {\eta}_2 \le 1$) \cite{24}, 
what would be the maximum value of ${\eta}_1$ (${\eta}_2$)? 

Applying cloning machine on one (say, on the first) party, it has been shown 
by Bandyopadhyay {\it et. al.} \cite{Bandyo} that ${\eta}_1 \le 1/3$. And, applying 
cloning machines locally and symmetrically ({\it i.e.}, 
${\eta}_1 = {\eta}_2 = \eta$) on both the particles, it has also been shown 
in \cite{Bandyo} that $\eta \le 1/{\sqrt{3}}$. Using most general type of 
local operations, Ghosh {\it et. al.} \cite{Ghosh} have shown that 
${\eta}_1 {\eta}_2 \le 1/3$. Zhou and Guo \cite{Zhang} have also obtained the
latter result by using the isotropy preserving 
(linear) map $\tilde{V} : \rho \mapsto {\rho}^{\prime}$, $\rho, {\rho}^{\prime}$ being 
density matrices of a qubit, where $\tilde{V}$ is defined as $\tilde{V}(I) = I$, 
$\tilde{V}({\sigma}_j) = \eta {\sigma}_j$ (for $j = 1, 2, 3$).

We denote here the Pauli spin matrices as 
$${\sigma}_1 = \left(
               \begin{array}{cc}
               0& 1\\
               1& 0
               \end{array}
               \right),~ {\sigma}_2 = \left(
               \begin{array}{cc}
               0& -i\\
               i& 0
               \end{array}
               \right),~ {\sigma}_3 = \left(
               \begin{array}{cc}
               1& 0\\
               0& -1
               \end{array}
               \right),$$   
expressed in the orthonormal basis $\{|0\rangle, |1\rangle\}$, where
$|0\rangle$ and $|1\rangle$ are eigen--states of ${\sigma}_3$ corresponding to
the eigen--values 1 and -1 respectively.

\section{Sufficient criterion for exact disentanglement}
Consider a set ${\cal S}_{commutative}$ of states of two two--level systems
possessed by 1
and 2, where the reduced density matrices of the particle 1 (say) commute with each other. It has been 
shown by Mor and Terno \cite{MT} that this set ${\cal S}_{commutative}$ can be exactly 
disentangled. Here we provide a simple proof of this fact, given in Ref. \cite{Ghosh3}.  

We can always take the set ${\cal S}_{commutative}$ as a subset of
$${\cal S}_{diagonal} = \{{\rho}_{12}(w) : Tr_2 [{\rho}_{12}(w)] = w P[|0\rangle_1] + (1 - w) P[|1\rangle_1], 0 \le w \le 1\},$$
where $\{|0\rangle_1, |1\rangle_1\}$ is an orthonormal basis in the two--dimensional Hilbert 
space of the party 1. Let Alice possess the two parties 1 and 3, 
while Bob holds the party 4, where the two parties 3 and 4 are connected by the 
classically correlated state $(1/2) (P[|00\rangle_{34}] + P[|11\rangle_{34}])$.
We consider this separable state as a teleportation channel for teleporting the states of the party 1 from the set ${\cal S}_{commutative}$. 
So the joint initial state of the three parties 1, 3 and 4 is 
$$(w P[|0\rangle_1] + (1 - w) P[|1\rangle_1]) \otimes (1/2) (P[|00\rangle_{34}] + P[|11\rangle_{34}]).$$ 
Alice first performs the projective measurement $\{P_1, P_2\}$ on her two parties 1 and 3, where $P_1 = P[|00\rangle_{13}] + P[|11\rangle_{13}]$ 
and $P_2 = P[|01\rangle_{13}] + P[|10\rangle_{13}]$. If $P_1$ clicks, Alice tells Bob (by a phone call) 
to do nothing, and so Bob's state then becomes $w P[|0\rangle_4] + (1 - w) P[|1\rangle_4]$ -- 
the state which Alice wants to teleport to Bob. On the other hand, if $P_2$
clicks, Alice tells Bob (by a phone call) to apply a unitary operation on his 
system which transforms $|0\rangle$ to $|1\rangle$ and $|1\rangle$ to $|0\rangle$ ({\it i.e.}, Bob has to operate the 
unitary operation ${\sigma}_x$ on his system, after being ensured that $P_2$
has been clicked in Alice's measurement). Again Bob's state becomes $w P[|0\rangle_4] + (1 - w) P[|1\rangle_4]$ -- 
the state which Alice wants to teleport to Bob. Thus exact teleportation of the party 1's state 
(from the set ${\cal S}_{commutative}$), from Alice to Bob, is complete. 
After this teleportation, each of the states from the set ${\cal S}_{commutative}$ (initially possessed 
by the parties 1 and 2) becomes exactly disentangled (now being possessed by the two parties 2 and 4), 
as no entanglement can be created by the LOCC operation used in the above--mentioned teleportation.

\section{Optimal disentanglement on the equatorial plane}
We discuss in this section the optimal disentanglement of a state of
two two--level systems shared between 1 and 2, taken at random from the set 
$${\cal S}_{equator} = \{{\rho}_{12} : Tr_2 [{\rho}_{12}] \in {\cal
P}_{equator}\},$$ 
where 
$${\cal P}_{equator} = \{w P[\alpha |0\rangle_1 + \beta |1\rangle_1] + (1 - w) P[\beta |0\rangle_1 -
\alpha |1\rangle_1] :$$
$$0 \le w \le 1~ {\rm and}~ \alpha, \beta \in I\!\!R ~{\rm with}~
\alpha^2 + \beta^2 = 1\},$$
{\it i.e.}, the set of all states of the party 1, each of whose Bloch vectors
lies on the equatorial plane. Here $\{|0\rangle_1, |1\rangle_1\}$ is an orthonormal basis of the Hilbert 
space of the party 1.  
In this direction, we shall first try to find out whether there is any realizable map $\tilde{V}$ which 
will isotropically shrink the Bloch vectors of the states of the set ${\cal
P}_{equator}$ by a factor $\lambda$ (where $-1 \le \lambda \le 1$), and then
we shall operate this map on the states of the party 1 to disentangle optimally the
states of the set ${\cal S}_{equator}$.

It has been shown in the Appendix that a (linear) map $\tilde{V}$ of the
above--mentioned type does exist, and is given by 
\begin{equation}
\label{appenmapV1}
\left.
\begin{array}{lcc}
\tilde{V} (I) &=& I,\\
\tilde{V} ({\sigma}_1) &=& \lambda {\sigma}_1,\\
\tilde{V} ({\sigma}_2) &=& (1 + \lambda) {\rm sin} {\theta} {\sigma}_1 + \{(1 +
\lambda) {\rm cos} {\theta} - \lambda\} {\sigma}_2\\
              & & - (1 - \lambda^2)^{1/2} {\rm sin} {\phi} {\sigma}_3\\
\tilde{V} ({\sigma}_3) &=& \lambda {\sigma}_3.
\end{array}
\right\}
\end{equation}

Next we apply this map on the states of the party 1 of the states ${\rho}_{12}$,
taken at random from the set ${\cal S}_{equator}$, in order to disentangle the
states ${\rho}_{12}$, optimally ({\it i.e.}, where $\lambda$ becomes maximum).
It has been shown in the Appendix that in this case of optimal disentanglement
of the states of the set ${\cal S}_{equator}$, we must have 
$${\rm sin} {\theta} = {\rm sin} {\phi} = 0, {\rm cos} {\theta} = \frac{1}{3},
\lambda = \frac{1}{2}.$$
It may be noted that under these optimality condition, the map $\tilde{V}$ of
equation (\ref{appenmapV1}) becomes physical.
 
\section{Disentanglement and teleportation}
From our discussions in the earlier sections, we got the following results
regarding disentanglement of entangled states of two
two--level systems possessed by 1 and 2, by applying the disentangling machines
on the states of the particle 2 (say) :\\
(i) In the optimal universal disentanglement process, an arbitrary density
matrix  
$${\rho}_{12} = \frac{1}{4} [I \otimes I + \vec{r}.\vec{\sigma} \otimes I + I
\otimes \vec{s}.\vec{\sigma} + \sum_{i, j = 1}^3 t_{ij} {\sigma}_i \otimes
{\sigma}_j ],$$
of the two parties 1 and 2, is transformed to the separable density matrix
$${\rho}^{\prime}_{12} = \frac{1}{4} [I \otimes I + \vec{r}.\vec{\sigma} \otimes I +
\lambda I
\otimes \vec{s}.\vec{\sigma} + \lambda \sum_{i, j = 1}^3 t_{ij} {\sigma}_i \otimes
{\sigma}_j ],$$
where the linear map $\tilde{V}$ (on the states of party 2), realizing this
process, is given by 
\begin{equation}
\label{universaldis}
\tilde{V} (I) = I,~ \tilde{V} ({\sigma}_1) = \lambda {\sigma}_1,~ \tilde{V} ({\sigma}_2) =
\lambda {\sigma}_2,~ \tilde{V} ({\sigma}_3) = \lambda {\sigma}_3,
\end{equation}
with $\lambda = 1/3$. 

(ii) In the optimal disentanglement process of all states of two parties 1 and
2, where each of the reduced density matrices of the party 2 lies on the 
equatorial plane, an arbitrary density matrix of the two parties 1 and 2
$${\rho}_{12} = \frac{1}{4} [I \otimes I + \vec{r}.\vec{\sigma} \otimes I + I
\otimes \vec{s}.\vec{\sigma} + \sum_{i, j = 1}^3 t_{ij} {\sigma}_i \otimes
{\sigma}_j ],$$
with $s_2 = 0$ (${\rho}_{12}$ being a representative of the corresponding class
of states),
is transformed to the separable density matrix
$${\rho}^{\prime}_{12} = \frac{1}{4} [I \otimes I + \vec{r}.\vec{\sigma} \otimes I + I
\otimes (\lambda s_1 {\sigma}_1 + \lambda s_3 {\sigma}_3)
+ \lambda \sum_{i = 1}^{3} \sum_{j \in \{1, 3\}} t_{ij} {\sigma}_i \otimes
{\sigma}_j + \mu \sum_{i = 1}^{3} t_{i2} {\sigma}_i \otimes
{\sigma}_2],$$
where the linear map $\tilde{V}$ (on the states of party 2), realizing this
process, is given by
\begin{equation}
\label{equatordis}
\tilde{V} (I) = I,~ \tilde{V} ({\sigma}_1) = \lambda {\sigma}_1,~ \tilde{V} ({\sigma}_2) =
\mu {\sigma}_2,~ \tilde{V} ({\sigma}_3) = \lambda {\sigma}_3,
\end{equation}
where $\lambda = 1/2, \mu = 0$. We call this optimal disentangling machine as 
the ``optimal equatorial disentangling machine."

(iii) In the exact disentanglement process of all entangled states of two
two--level systems possessed by 1 and 2, where the reduced density matrices of 
party 2 commute with each other, an arbitrary density matrix
$${\rho}_{12} = \frac{1}{4} [I \otimes I + \vec{r}.\vec{\sigma} \otimes I + I
\otimes \vec{s}.\vec{\sigma} + \sum_{i, j = 1}^3 t_{ij} {\sigma}_i \otimes
{\sigma}_j ],$$
with $s_1 = s_2 = 0$ (${\rho}_{12}$ being a representative of the corresponding class of states),
is transformed to the separable density matrix
$${\rho}^{\prime}_{12} = \frac{1}{4} [I \otimes I + \vec{r}.\vec{\sigma}
\otimes I + \lambda I \otimes s_3 {\sigma}_3
+ \sum_{i = 1}^{3} {\sigma}_i \otimes (t_{i1} \nu {\sigma}_1 + t_{i2} \mu
{\sigma}_2 + t_{i3} \lambda {\sigma}_3)],$$
where the linear map $\tilde{V}$ (on the states of party 2), realizing this
process, is given by
\begin{equation}
\label{diametredis}
\tilde{V} (I) = I,~ \tilde{V} ({\sigma}_1) = \nu {\sigma}_1,~ \tilde{V} ({\sigma}_2) =
\mu {\sigma}_2,~ \tilde{V} ({\sigma}_3) = \lambda {\sigma}_3,
\end{equation}
where $\lambda = 1, \nu = \mu = 0$. We call this disentangling machine 
as the ``exact disentangling machine" \cite{whyitisexact}.    

In this section we shall show that each of the maps $\tilde{V}$ in equations 
(\ref{universaldis}) -- (\ref{diametredis}) (together with their 
generalization), can be realized by sending the states of party 2 through
some separable channels, each of which is a specific type of Bell mixture. For 
this purpose, we take here the teleportation protocol of Bennett {\it et al}. 
\cite{BBCJPW}, which exactly teleports an arbitrary qubit through the channel 
$|{\psi}^{+}\rangle = (1/{\sqrt{2}}) (|01\rangle + |10\rangle)$, {\it 
irrespective of whatever the channel may be}. We designate this protocol by 
$B^{|{\psi}^+\rangle}$.

Consider now the teleportation of an arbitrary state of a party 1 through the
channel 
\begin{equation}
\label{channel}
{\rho}^{Bell}_{23} = w_1 P[|{\psi}^{+}\rangle_{23}] + w_2
P[|{\psi}^{-}\rangle_{23}] + w_3 P[|{\phi}^{+}\rangle_{23}] + w_4
P[|{\phi}^{-}\rangle_{23}], 
\end{equation}
between the two parties 2 and 3 to party 3, using $B^{|{\psi}^+\rangle}$, where
$0 \le w_1, w_2, w_3, w_4 \le 1$ and $\sum_{i=1}^{4} w_i = 1$, and where
$|{\psi}^{\pm}\rangle_{23} = (1/{\sqrt{2}}) (|01\rangle_{23} \pm
|10\rangle_{23})$, $|{\phi}^{\pm}\rangle_{23} = (1/{\sqrt{2}}) (|00\rangle_{23} \pm
|11\rangle_{23})$. Thus
an arbitrary state $\alpha |0\rangle_1 + \beta |1\rangle_1$ of party 1 will 
surface as the following state of party 3 :
$$w_1 P[\alpha |0\rangle_3 + \beta |1\rangle_3] + w_2 P[\alpha |0\rangle_3 -
\beta |1\rangle_3] + w_3 P[\alpha |1\rangle_3 + \beta |0\rangle_3] + w_4
P[\alpha |1\rangle_3 - \beta |0\rangle_3].$$
So the linear map $\tilde{V}$ (from the states of party 1 to the states of 
party 3) is given by 
$$\tilde{V} (I) = I,$$
$$\tilde{V} ({\sigma}_1) = (w_1 - w_2 + w_3 - w_4) {\sigma}_1,$$
$$\tilde{V} ({\sigma}_2) = (w_1 - w_2 - w_3 + w_4) {\sigma}_2,$$
$$\tilde{V} ({\sigma}_3) = (w_1 + w_2 - w_3 - w_4) {\sigma}_3.$$

It can be easily seen that :\\
(i) the optimal universal disentangling machine ({\it i.e.}, the map
$\tilde{V}$ of equation (\ref{universaldis}), except a possible change in 
the parties) can be obtained by teleporting (using the protocol
$B^{|{\psi}^+\rangle}$) the states of one party (say, party 2)
through the separable Werner channel \cite{Werner}   
$${\rho}^{Werner}_{23} = \frac{1}{3} P[|{\psi}^{+}\rangle_{23}] + \frac{1}{6}
I_2 \otimes I_3,$$
from 2 to 3 (as here $\nu = \mu = \lambda = 1/3$);

(ii) the ``optimal equatorial disentangling machine" ({\it i.e.}, the map
$\tilde{V}$ of equation (\ref{equatordis}), except a possible change in
the parties) can be obtained by teleporting (using the protocol
$B^{|{\psi}^+\rangle}$) the states of one party (say, party 2) through the
separable channel
$$\frac{1}{2} P[|{\psi}^{+}\rangle_{23}] + \frac{1}{4}
P[|{\psi}^{-}\rangle_{23}] + \frac{1}{4} P[|{\phi}^{+}\rangle_{23}],$$
from $2$ to $3$ (as here $\nu = \lambda = 1/2$ and $\mu =
0$);
 
(iii) the ``exact disentangling machine" ({\it i.e.}, the map
$\tilde{V}$ of equation (\ref{diametredis}), except a possible change in
the parties) can be obtained by teleporting (using the protocol
$B^{|{\psi}^+\rangle}$) the states of one party (say, party 2) through the
separable channel
$$\frac{1}{2} P[|{\psi}^{+}\rangle_{23}] + \frac{1}{2}
P[|{\psi}^{-}\rangle_{23}],$$
from $2$ to $3$ (as here $\nu = \mu = 0$ and $\lambda =
1$). 
 
In general, we have established the following result :\\
Teleportation (using the protocol
$B^{|{\psi}^{+}\rangle}$) of the states of party 2 of any density matrix
${\rho}_{12}$ of two two--level systems 1 and 2 through the channel
${\rho}^{Bell}_{AB} = w_1 P[|{\psi}^{+}\rangle_{AB}] + w_2
P[|{\psi}^{-}\rangle_{AB}] + w_3 P[|{\phi}^{+}\rangle_{AB}] + w_4
P[|{\phi}^{-}\rangle_{AB}]$ (where $0 \le w_1, w_2, w_3, w_4 \le 1$, and $\sum_{i =
1}^{4} w_i = 1$), from $A$ to $B$, gives rise to the physical linear map 
$\tilde{V}$ (from the states of party 2 to those of party $B$), where
\begin{equation}
\label{bellmix}
\left.
\begin{array}{lcc}
\tilde{V} (I) &=& I,\\
\tilde{V} ({\sigma}_1) &=& \lambda_1 {\sigma}_1,\\
\tilde{V} ({\sigma}_2) &=& \lambda_2 {\sigma}_2,\\
\tilde{V} ({\sigma}_3) &=& \lambda_3 {\sigma}_3,
\end{array}
\right\}
\end{equation}  
with the following one--to--one correspondence :
\begin{equation}
\label{numulambda}
\lambda_1 = w_1 - w_2 + w_3 - w_4,~ \lambda_2 = w_1 - w_2 - w_3 + w_4,~ \lambda_3 = w_1 + w_2
- w_3 - w_4,
\end{equation}
and
$$w_1 = (1 + {\lambda}_1 + {\lambda}_2 + {\lambda}_3)/4,~ w_2 = (1 - {\lambda}_1 - {\lambda}_2 + {\lambda}_3)/4,$$
\begin{equation}
\label{w1234}
w_3 = (1 + {\lambda}_1 - {\lambda}_2 - {\lambda}_3)/4,~ w_4 = (1 - {\lambda}_1 + {\lambda}_2 - {\lambda}_3)/4,
\end{equation}
and where
\begin{equation}
\label{modvalue}
0 \le |\lambda_1|,~ |\lambda_2|,~ |\lambda_3| \le 1.
\end{equation}

Does the converse of this result also hold good?  In other words, 
if $\tilde{V}$ is a physical map (from density matrices to density matrices,
with a possible change in the parties) having the property that, $\tilde{V} (I)
= I, \tilde{V} ({\sigma}_j) = {\lambda}_j {\sigma}_j$, ${\lambda}_j \in I\!\!R$
with $0 \le |{\lambda}_j| \le 1 (j = 1, 2, 3)$, would it then be possible to
get this map (in an one--to--one manner) by teleporting the density matrices of
a party through the channel ${\rho}^{Bell}_{AB}$ (by a suitable choice of the 
$w_i$'s), using the protocol $B^{|{\psi}^{+}\rangle}$? We shall now show that
this is indeed true. 

{\noindent {\bf Lemma} : If $\tilde{V}$ is a physical map 
(from the density matrices of a party $A$ to the density matrices of a party
$B$, where $A$ and $B$ may not be the same), having the property that $\tilde{V} (I)
= I, \tilde{V} ({\sigma}_j) = {\lambda}_j {\sigma}_j$, ${\lambda}_j \in I\!\!R$
with $0 \le |{\lambda}_j| \le 1 (j = 1, 2, 3)$, the four quantities $w_1 = (1 +
{\lambda}_1 + {\lambda}_2 + {\lambda}_3)/4, w_2 = (1 -
{\lambda}_1 - {\lambda}_2 + {\lambda}_3)/4, w_3 = (1 +
{\lambda}_1 - {\lambda}_2 - {\lambda}_3)/4, w_4 = (1 -
{\lambda}_1 + {\lambda}_2 - {\lambda}_3)/4,$ will then satisfy the properties $0 \le
w_1, w_2, w_3, w_4 \le 1$ and $\sum_{i = 1}^{4} w_i = 1.$}

{\noindent {\bf Proof} : {\underline {Case I :}} Consider $A$ and $B$ to be the
same.\\
The unitary operator $U$, acting on the joint Hilbert space 
${\cal H}_A \otimes {\cal H}_B \otimes {\cal H}_C$ of the parties $A, B$, and some ancilla $C$ 
(the initial state of the joint system ${\cal H}_B \otimes {\cal H}_C$ is taken as $|M\rangle_{BC}$), 
realizing the map $\tilde{V}$, is given as follows :
\begin{equation}
\label{UofV}
\left.
\begin{array}{lcr}
U(|0\rangle_A \otimes |M\rangle_{BC}) &=& a_0 |0\rangle_A \otimes |M_0\rangle_{BC} + b_0 |1\rangle_A \otimes |M_1\rangle_{BC},\\
U(|1\rangle_A \otimes |M\rangle_{BC}) &=& b_0 |0\rangle_A \otimes |\tilde{M_0}\rangle_{BC} + a_0 |1\rangle_A \otimes |\tilde{M_1}\rangle_{BC},\\
\end{array}
\right\}
\end{equation}
where $\{|0\rangle_A, |1\rangle_A\}$ is an orthonormal basis in ${\cal H}_A$, $a_0, b_0$ are non--negative numbers (with $a_0^2 + b_0^2 = 1$), $|M\rangle, |M_0\rangle, |M_1\rangle, |\tilde{M_0}\rangle, |\tilde{M_1}\rangle$ are normalized states in ${\cal H}_B \otimes {\cal H}_C$, and
$$ {\lambda}_3 = 2a_0^2 - 1 = 1 - 2b_0^2,~ {\lambda}_1 = a_0^2 \Re \{\langle M_0|\tilde{M_1} \rangle\} + b_0^2 \Re \{\langle \tilde{M_0}|M_1 \rangle\},$$
\begin{equation}
\label{lambdasa0b0}
{\lambda}_2 = a_0^2 \Re \{\langle M_0|\tilde{M_1} \rangle\} - b_0^2 \Re \{\langle \tilde{M_0}|M_1 \rangle\},
\end{equation}
\begin{equation}
\langle M_1|M_0 \rangle = \langle M_1|\tilde{M_1} \rangle = \langle \tilde{M_1}|\tilde{M_0} \rangle = \langle M_0|\tilde{M_0} \rangle = 0,       
\end{equation} 
\begin{equation}
\Im \{\langle M_0|\tilde{M_1} \rangle\} = \Im \{\langle \tilde{M_0}|M_1 \rangle \} = 0.
\end{equation}
Using the (given) definitions of $w_i$'s ($i = 1, 2, 3, 4$), we see that
\begin{equation}
\label{wandmachine}
w_1 = a_0^2 {\rm cos}^2 \theta,~ w_2 = a_0^2 {\rm sin}^2 \theta,~ w_3 = b_0^2 {\rm cos}^2 \phi,~ w_4 = b_0^2 {\rm sin}^2 \phi,
\end{equation}
where we have taken $\Re \{\langle M_0|\tilde{M_1} \rangle\} = {\rm cos} 2\theta$ and $\Re \{\langle \tilde{M_0}|M_1 \rangle\} = {\rm cos} 2\phi$. 
And using equation (\ref{wandmachine}), the fact that $0 \le a_0, b_0 \le 1$, and the fact that $\theta, \phi \in I\!\!R$, one can easily see that 
$0 \le w_1, w_2, w_3, w_4 \le 1$ and $\sum_{i = 1}^{4} w_i = 1$.\\ 
{\underline {Case II :}} Consider $A$ and $B$ to be different.\\
The proof is same as that in Case I, except for an interchange between $A, B$ on
the right hand sides of equation (\ref{UofV}). ${\Box}$ }

The channel ${\rho}^{Bell}_{AB} = w_1 P[|{\psi}^{+}\rangle_{AB}] + w_2
P[|{\psi}^{-}\rangle_{AB}] + w_3 P[|{\phi}^{+}\rangle_{AB}] + w_4
P[|{\phi}^{-}\rangle_{AB}]$ will be separable iff $0 \le w_i \le 1/2$, for $i =
1, 2, 3, 4$ \cite{BenPRA}. Thus if ${\cal S}$ is any set of entangled states
${\rho}_{12}$ of two two--level systems 1 and 2, each state ${\rho}_{12}$ of ${\cal S}$ will
become a separable (or, {\it unentangled}) state ${\rho}^{\prime}_{1B}$ when we teleport the state
of the party 2 through this channel, from $A$ to $B$, whatever be the protocol
we use. But in the following theorem (the main result of this section), we
shall use the above--mentioned Lemma, and hence the protocol
$B^{|{\psi}^{+}\rangle}$.   
         
{\noindent {\bf Theorem} : Consider the set ${\cal S}_{universal}$ of all entangled
states ${\rho}_{12}$ of two two--level systems 1 and 2. If $\tilde{V}$
is a physical map, having the property that $\tilde{V} (I) = I,
\tilde{V} ({\sigma}_j) = {\lambda}_j {\sigma}_j, 0 \le
{\lambda}_j \le 1~ ({\rm for}~ j = 1, 2, 3)$, this map will then unentangle 
each member of ${\cal S}_{universal}$, by acting on the states of the party 2, if and only if 
$\sum_{j = 1}^{3} {\lambda}_j \le 1$.}

{\noindent {\bf Proof} : {\underline{Sufficient part :}} Using Lemma 1, we can 
uniquely find the coefficients
$w_i$ ($i = 1, 2, 3, 4$) of the channel ${\rho}^{Bell}_{AB} = w_1
P[|{\psi}^{+}\rangle_{AB}] + w_2
P[|{\psi}^{-}\rangle_{AB}] + w_3 P[|{\phi}^{+}\rangle_{AB}] + w_4
P[|{\phi}^{-}\rangle_{AB}]$. As each of the ${\lambda}$'s is
non--negative, therefore, $w_1$ is the highest coefficient. And from
the condition that $\sum_{j = 1}^{3}
{\lambda}_j \le 1$, we see that ${\rm max} \{w_1, w_2, w_3, w_4\} = w_1 \le
1/2$. Hence the channel ${\rho}^{Bell}_{AB}$ must be separable. So, the 
unentanglement in question, can be achieved by teleporting (using the protocol 
$B^{|{\psi}^{+}\rangle}$) the states of party 2 through a seperable 
channel ${\rho}^{Bell}_{AB} = w_1 P[|{\psi}^{+}\rangle_{AB}] + w_2
P[|{\psi}^{-}\rangle_{AB}] + w_3 P[|{\phi}^{+}\rangle_{AB}] + w_4
P[|{\phi}^{-}\rangle_{AB}]$, from $A$ to $B$. 

{\underline{Necessary part :}} We shall show here that if 
$\sum_{j = 1}^{3} {\lambda}_j > 1$, there exist(s) some state(s) which can not 
be unentangled by the given map $\tilde{V}$.\\
Consider the state ${\rho}_{12} = P[(1/{\sqrt{2}}) (|0\rangle_1 \otimes |0\rangle_2 + |1\rangle_1 \otimes |1\rangle_2)]$ 
of two parties 1 and 2. Applying the above map $\tilde{V}$ on the states of
the party 2, one gets the following density matrix 
$${\rho}^{\prime}_{12} = \frac{1}{4} \left[
                                     \begin{array}{cccc}
                                     1 + {\lambda}_3 &0& 0& {\lambda}_1 + {\lambda}_2\\
0 &1 - {\lambda}_3& {\lambda}_1 - {\lambda}_2& 0\\
0 &{\lambda}_1 - {\lambda}_2& 1 - {\lambda}_3& 0\\ 
{\lambda}_1 + {\lambda}_2 &0& 0& 1 + {\lambda}_3
\end{array}
\right]$$
Using Peres--Horodecki theorem \cite{PeresHoro}, one can easily see that the density matrix ${\rho}^{\prime}_{12}$ will be entangled if 
$\sum_{j = 1}^{3} {\lambda}_j > 1$ and $0 \le {\lambda}_1, {\lambda}_2, {\lambda}_3
\le 1$. Thus we see that if the map $\tilde{V}$ unentangles each member of ${\cal S}$ (by
acting on the states of the party 2), we must have $\sum_{j = 1}^{3}
{\lambda}_j \le 1$. $\Box$}            

Here it is to be noted that, the notion of `unentanglement' is a generalization
of the notion of `disentanglement', because a physical map $\tilde{V}$, defined as
$\tilde{V} (I) = I, \tilde{V} ({\sigma}_j) = {\lambda}_j {\sigma}_j$ ($j= 1, 2,
3$), which unentangles each state ${\rho}_{12}$ of two two--level systems 1 and 2, after
acting on one party (say, party 2) of the state, disentangles the three sets 
${\cal S}_j = \{{\rho}_{12} | Tr_1 [{\rho}_{12}] = (1/2) [I + s_j {\sigma}_j], {\rm where} |s_j| \le 1\}$ $(j
= 1, 2, 3)$, separately. So this unentanglement will be (i) a `universal
disentanglement' iff ${\lambda}_1 = {\lambda}_2 = {\lambda}_3$, (ii) an
`equatorial disentanglement' iff two and only two of the ${\lambda}_i$'s are
equal, (iii) a `exact disentanglement' iff two of the ${\lambda}_i$'s are
equal to 0 and the third is equal to 1.

\section{Conclusion}  
We obtained in this paper the best disentangling machine of the set of all states 
of two two--level systems, where the states of one party lie on a single great
disc, 
the corresponding map being $\tilde{V} (I) = I, \tilde{V} ({\sigma}_1) =
(1/2) ({\sigma}_1), \tilde{V} ({\sigma}_2) = 0, \tilde{V} ({\sigma}_3) =
(1/2) ({\sigma}_3)$). One may now ask that if the reduced density matrices of
both the parties lie on two great discs, and if we operate two disentangling
machines separately on the states of these two parties, can we obtain a better
disentanglement (e.g., using the map $\tilde{V} (I) = I, \tilde{V}
({\sigma}_1) = (1/2 + \epsilon) ({\sigma}_1), \tilde{V} ({\sigma}_2) = 0,
\tilde{V} ({\sigma}_3) = (1/2 + \delta) ({\sigma}_3)$, on the states of both
parties, $\epsilon$ and $\delta$ being any two positive numbers)? The answer is
no, because whenever we take the shrinking factor
$\lambda$ to be greater than $1/2$, the corresponding map becomes unphysical 
in this case. So the best disentangling machine in this case, can be obtained 
by using the machine on one party only.

The fact that a linear map $\tilde{V}$, defined as
$\tilde{V} (I) = I, \tilde{V} ({\sigma}_j) = \sum_{k = 1}^{3} {\lambda}_{jk}
{\sigma}_k$ (${\lambda}_{jk} \in I\!\!R$, and $j, k = 1, 2, 3$), can not
decrease entanglement of any state ${\rho}_{12}$ if $\sum_{k = 1}^{3}
{\lambda}_{jk}^2 = 1$ for all $k$ (the case when $\sum_{k = 1}^{3}
{\lambda}_{jk}^2 > 1$ is discarded as then the map $\tilde{V}$ becomes
unphysical), for disentanglement of ${\rho}_{12}$ (using the machine
on one side only), we must have $\sum_{k = 1}^{3}
{\lambda}_{jk}^2 < 1$ for some or all $j = 1, 2, 3$. And this we have observed
in the cases of optimal universal disentanglement, optimal equatorial 
disentanglement, or exact disentanglement (related to a commuting set). We 
obtained these optimal machines by teleporting the states of one side through 
some separable channels. And this has been shown in two steps : firstly we have 
established an one--to--one correspondence between the physical map  
$\tilde{V} (I) = I, \tilde{V} ({\sigma}_j) = {\lambda}_j ({\sigma}_j)$, 
($j = 1, 2, 3$), and teleportation of qubits through a mixture of Bell states,
and secondly we have shown that, each of the
processes in which ${\rho}_{12}$ becomes separable (by using the map on one
side, where $0 \le {\lambda}_1, {\lambda}_2, {\lambda}_3 \le 1$), can be
achieved by this one--to--one correspondence if and only if $\sum_{j = 1}^{3}
{\lambda}_j \le 1$.      

\vspace{0.4cm}
{\noindent {\bf Acknowledgement} :} 

U.S. thanks Dipankar Home for encouragement and acknowledges partial support by
the Council of Scientific and Industrial Research, Government of India, New
Delhi. 

\vspace{0.5cm}
\begin{center}
{\large {\bf Appendix}}
\end{center}

We discuss here, in details, the optimal disentanglement of a state of
two two--level systems shared between 1 and 2, taken at random from the set 
$${\cal S}_{equator} = \{{\rho}_{12} : Tr_2 [{\rho}_{12}] \in {\cal
P}_{equator}\},$$ 
where 
$${\cal P}_{equator} = \{w P[\alpha |0\rangle_1 + \beta |1\rangle_1] + (1 - w) P[\beta |0\rangle_1 -
\alpha |1\rangle_1] :$$
$$0 \le w \le 1~ {\rm and}~ \alpha, \beta \in I\!\!R ~{\rm with}~
\alpha^2 + \beta^2 = 1\},$$
{\it i.e.}, the set of all states of the party 1, each of whose Bloch vectors
lies on the equatorial plane. Here $\{|0\rangle_1, |1\rangle_1\}$ is an orthonormal basis of the Hilbert 
space of the party 1.  

We first find out a realizable map $\tilde{V}$ which 
will isotropically shrink the Bloch vectors of the states of the set ${\cal
P}_{equator}$ by a factor $\lambda$ (where $-1 \le \lambda \le 1$), and then
we operate this map on the states of the party 1 to disentangle optimally the
states of the set ${\cal S}_{equator}$.

Consider any equatorial (pure) state 
$$|\psi\rangle = \alpha |0\rangle + \beta |1 \rangle,$$ 
so that $\alpha, \beta \in I\!\!R$ with $\alpha^2 + \beta^2 = 1$. And let $U$ be a unitary operator (acting on the joint
Hilbert space of the equatorial states and states of some ancilla) realizing the
map $\tilde{V}$. As we want to maintain the isotropies of the basis vectors 
$|0\rangle , |1\rangle$, $U$ must behave as follows :
$$
\left.
\begin{array}{lcr}
U|0M\rangle &=& m_{0} |0M_{0}\rangle + m^{\prime}_{0} |1{\overline{M_0}}\rangle,\\
U|1M\rangle &=& m_{1} |0M_{1}\rangle + m^{\prime}_{1} |1{\overline{M_1}}\rangle, 
\end{array}
\right\}
\eqno{(A1)}$$
where $|M\rangle$ is the input ancilla state and $|M_{0}\rangle, |{\overline{M_0}},
|M_{1}\rangle, |{\overline{M_1}}\rangle$ are the output ancilla states with 
$\langle{M_0}|{\overline{M_0}}\rangle = 0 = \langle{M_1}|{\overline{M_1}}\rangle$
and $m_0, m_1, {m^{\prime}}_0, {m^{\prime}}_1 \ge 0$.
Unitarity of the operation imply
$$
m_0^2 +  {m^{\prime}}_0^2 = m_1^2 + {m^{\prime}}_1^2 = 1
\eqno{(A2)}$$
and 
$$
m_0 m_1 \langle{M_0}|{M_1}\rangle + {m^{\prime}}_0 {m^{\prime}}_1 \langle{\overline{M_0}}|{\overline{M_1}}\rangle = 0. 
\eqno{(A3)}$$
Therefore, according to our demand (of maintaing isotropies of the equatorial 
states),
$$\tilde{V} (P[|\psi\rangle]) = \tilde{V} (\frac{1}{2} [I + 2 \alpha \beta {\sigma}_1 + 
(\alpha^2 - \beta^2) {\sigma}_3]) \doteq Tr_M [P[U(| \psi M \rangle)]] =$$ 
$$\frac{1}{2} [I + 2 \alpha \beta \Re \{m_0
{m^{\prime}}_1 \langle{\overline{M_1}}|{M_0}\rangle + m_1 {m^{\prime}}_0 \langle{\overline{M_0}}|{M_1}\rangle \} {\sigma}_1$$ 
$$- 2 \alpha \beta \Im \{ \ m_0 {m^{\prime}}_1 \langle {\overline{M_1}}|{M_0}\rangle + 
m_1 {m^{\prime}}_0 \langle {\overline{M_0}}|{M_1}\rangle \} {\sigma}_2$$
$$+ ({\alpha}^2 (m_0^2 - {m^{\prime}}_0^2) + {\beta}^2 (m_1^2 - {m^{\prime}}_1^2) + 
2 \alpha \beta \Re \{m_0 m_1 \langle {M_0}|{M_1} \rangle - {m^{\prime}}_0 {m^{\prime}}_1 
\langle {\overline{M_0}}|{\overline{M_1}} \rangle\}) {\sigma}_3]$$
$$
\equiv \frac{1}{2} [I + 2 \lambda \alpha \beta {\sigma}_1 + \lambda (\alpha^2 -
\beta^2) {\sigma}_3],
\eqno{(A4)}$$
where $\lambda \in I\!\!R$, with $|\lambda| \le 1$. $\lambda$ is called
the {\it shrinking factor}.

Here equation (A4) has to be satisfied for all $\alpha, \beta \in
I\!\!R$, where the map $\tilde{V}$ is assumed to be linear.
So we must have\footnote{Here $\Re{\{z\}}$ and $\Im{\{z\}}$ denote the real and
imaginary parts of $z$ respectively.}
$$
\left.
\begin{array}{ccc}
\Re \{m_0 m^{\prime}_1 \langle \overline{M_1}|M_0 \rangle + m_1 m^{\prime}_0
\langle \overline{M_0}|M_1 \rangle\} &=& \lambda,\\
\Im \{m_0 m^{\prime}_1 \langle \overline{M_1}|M_0 \rangle + m_1 m^{\prime}_0
\langle \overline{M_0}|M_1 \rangle\} &=& 0,\\
\Re \{m_0 m_1 \langle M_0|M_1 \rangle - m^{\prime}_0 m^{\prime}_1 
\langle \overline{M_0}|\overline{M_1} \rangle\} &=& 0,\\
{m_0}^2 - {m^{\prime}_0}^2 &=& \lambda,\\
{m_1}^2 - {m^{\prime}_1}^2 &=& -\lambda.
\end{array}
\right\}
\eqno{(A5)}$$
Using equations (A2) -- (A5), we get
$$
\left.
\begin{array}{llcrr}
m_0 &=& m^{\prime}_1 &=& (\frac{1 + \lambda}{2})^{1/2},\\
m_1 &=& m^{\prime}_0 &=& (\frac{1 - \lambda}{2})^{1/2}.
\end{array}
\right\}
\eqno{(A6)}$$
As not each pair of states on the equatorial plane ${\cal P}_{equator}$
commute, therefore the states of the set ${\cal S}_{equator}$ can not be
exactly disentangled by this map $\tilde{V}$ \cite{Ghosh3}. So $|\lambda| < 1$.
Thus we see from equation (A6) that $m_0, m^{\prime}_0, m_1,
m^{\prime}_1 > 0$. Using this fact and equations (A3) --
(A6), we have
$$
\left.
\begin{array}{ccc}
\Re \{\langle M_0|M_1 \rangle\} &=& 0,\\
\Re \{\langle \overline{M_0}|\overline{M_1} \rangle\} &=& 0,\\
\Im \{\langle M_0|M_1 \rangle\} - \Im \{\langle
\overline{M_0}|\overline{M_1} \rangle\} &=& 0,\\ 
(1 + \lambda) \Im \{\langle \overline{M_1}|M_0 \rangle\} + (1 - \lambda) \Im
\{\langle \overline{M_0}|M_1 \rangle\} &=& 0,\\
(1 + \lambda) \Re \{\langle \overline{M_1}|M_0 \rangle\} + (1 - \lambda) \Re
\{\langle \overline{M_0}|M_1 \rangle\} &=& 2\lambda.
\end{array}
\right\}
\eqno{(A7)}$$
Thus the linear map $\tilde{V}$ works (on all density matrices) as
follows\footnote{The mapping of $\tilde{V}$ on ${\sigma}_2 \equiv
P[(1/{\sqrt{2}}) (|0\rangle + i|1\rangle)] - P[(1/{\sqrt{2}}) (|0\rangle -
i|1\rangle)]$ is obtained by considering the actions of the unitary operator
$U$ of equation (A1) on the states $(1/{\sqrt{2}}) (|0\rangle \pm i|1\rangle)
\otimes |M\rangle$, and then taking traces over the machine states.} :
$$
\left.
\begin{array}{lcc}
\tilde{V} (I) &=& I,\\
\tilde{V} ({\sigma}_1) &=& \lambda {\sigma}_1,\\
\tilde{V} ({\sigma}_2) &=& (1 + \lambda) \Im \{\langle \overline{M_1}|M_0
\rangle\} {\sigma}_1+ \{(1 + \lambda) \Re \{\langle \overline{M_1}|M_0
\rangle\} - \lambda\} {\sigma}_2\\
              & & - (1 - \lambda^2)^{1/2} \Im \{\langle M_0|M_1 \rangle\} {\sigma}_3\\
\tilde{V} ({\sigma}_3) &=& \lambda {\sigma}_3.
\end{array}
\right\}
\eqno{(A8)}$$
The map $\tilde{V}$ (in equation (A8)) has to satisfy the following
conditions in order that it will be a physical map, {\it i.e.}, it will map every density matrix
$\rho = (1/2) [I + \vec{r}.\vec{\sigma}]$ (where $|\vec{r}|^2 \equiv r_1^2 + r_2^2 + r_3^2 \le 1$) to a density matrix $\rho^{\prime}$ : 
$$\lambda^2 |\vec{r}|^2 + 2\lambda\{(1 +
\lambda)\Im\{\langle \overline{M_1}|M_0 \rangle\}r_1 - (1 -
\lambda^2)^{1/2}\Im\{\langle M_0|M_1 \rangle\}r_3\}r_2 +$$
$$
[(1 + \lambda)^2 |\langle \overline{M_1}|M_0 \rangle|^2 + (1 -
\lambda^2)(\Im\{\langle M_0|M_1 \rangle\})^2 - 2\lambda(1 +
\lambda)\Re\{\langle \overline{M_1}|M_0 \rangle\}]r_2^2 \le 1.
\eqno{(A9)}$$
   
Our next task is to disentangle the states taken at random from the set 
$${\cal S}_{equator} =$$
$$\{\rho_{12} : Tr_2 [\rho_{12}] = w
P[\alpha |0\rangle_1 + \beta |1\rangle_1] + (1 - w) P[\beta |0\rangle_1 -
\alpha |1\rangle_1],$$
$${\rm where}~ 0 \le w \le 1~ {\rm and}~ \alpha, \beta \in I\!\!R ~{\rm with}~
\alpha^2 + \beta^2 = 1\},$$
by 
applying the linear map\footnote{So there are of course some definite relations
among $\lambda, l, m, n$.} $\tilde{T}$,
defined as 
$$
\left.
\begin{array}{ccc}
\tilde{T} (I) &=& I,\\
\tilde{T} (\sigma_1) &=& \lambda \sigma_1,\\
\tilde{T} (\sigma_2) &=& l \sigma_2 + m \sigma_1 + n \sigma_3,\\ 
\tilde{T} (\sigma_3) &=& \lambda \sigma_3
\end{array}
\right\}
\eqno{(A10)}$$
(where $\lambda, l, m, n$ are real numbers, with $|\lambda| < 1$), on the states of 
the party 1. 
Obviously, the map $\tilde{T}$ (of equation (A10)) is a
generalization of the map $\tilde{V}$ (of equation (A8)). Now any
pure state $|\psi\rangle_{12}$ of the set ${\cal S}_{equator}$ can be taken in the following Schmidt form :
$$
|\psi\rangle_{12} = a|0^{\prime}\rangle_1 \otimes |0^{\prime}\rangle_2 + 
b|1^{\prime}\rangle_1 \otimes |1^{\prime}\rangle_2, 
\eqno{(A11)}$$
where $\{|0^{\prime}\rangle_j, |1^{\prime}\rangle_j\}$ is an orthonormal basis
in the two dimensional Hilbert space of the party $j$ $(j= 1, 2)$, $a, b$ are
non--negative numbers with $a^2 + b^2 = 1$, and $|0^{\prime}\rangle_1 = \alpha
|0\rangle_1 + \beta |1\rangle_1, |1^{\prime}\rangle_1 = \beta |0\rangle_1 - \alpha
|1\rangle_1$, where $\alpha, \beta$ are real numbers with $\alpha^2 + \beta^2 =
1$ ({\it i.e.}, they are equatorial states). As the action of the map
$\tilde{T}$ does not depend on the choice of the factors $\alpha, \beta$, and
as no operation has to be done on the states of the party 2, therefore, 
without loss of generality, we can take $|\psi\rangle_{12}$ as :
$$
|\psi\rangle_{12} = a|0\rangle_1 \otimes |0\rangle_2 + b|1\rangle_1 \otimes |1\rangle_2. 
\eqno{(A12)}$$
Now, in the ordered orthonormal basis $\{|0\rangle_1 \otimes |0\rangle_2,
|0\rangle_1 \otimes |1\rangle_2, |1\rangle_1 \otimes |0\rangle_2, |1\rangle
\otimes |1\rangle_2\}$, $(\tilde{T} \otimes I)
(P[|\psi\rangle_{12}])$ $(= {\rho}^{\prime})$ is given by :
$${\rho}^{\prime} =$$
$$
\frac{1}{4} \times 
\left[
\begin{array}{cccc} 
(1 + \lambda)2a^2 &2inab& 0& 2ab(\lambda + l + im)\\
-2inab &(1 - \lambda)2b^2& 2ab(\lambda - l - im)& 0\\
0 &2ab(\lambda - l + im)& (1 - \lambda)2a^2& -2inab\\ 
2ab(\lambda + l - im) &0& 2inab& (1 + \lambda)2b^2
\end{array}
\right]
\eqno{(A13)}$$
Here ${\rho}^{\prime}$ is a self--adjoint operator of trace 1, for every choice
of $a, b$. In order that the map $\tilde{T}$ will be a physical map, the
mapping of $\tilde{T} \otimes I$ over each state
$|\psi\rangle_{12}$ should produce a density operator (of two two--level
systems) -- the conditions for
which are 
$$
\left.
\begin{array}{ccc}
1 - {\lambda}^2 - n^2 &\ge& 0,\\
(1 - l^2 - m^2 - n^2) - \lambda(1 + l^2 + m^2 - 2l - n^2) - 2{\lambda}^2(1 - l)
&\ge& 0,\\
(1 - l^2 - m^2 - n^2)^2 - 4{\lambda}^2(1 - l)^2 &\ge& 0.
\end{array}
\right\}
\eqno{(A14)}$$
And the conditions for disentanglement of $|\psi\rangle_{12}$ are given by
(using the Peres--Horodecki theorem \cite{PeresHoro}),
$$
\left.
\begin{array}{ccc}
1 - {\lambda}^2 - n^2 &\ge& 0,\\
(1 - l^2 - m^2 - n^2) - \lambda(1 + l^2 + m^2 + 2l - n^2) - 2{\lambda}^2(1 + l)
&\ge& 0,\\
(1 - l^2 - m^2 - n^2)^2 - 4{\lambda}^2(1 + l)^2 &\ge& 0.
\end{array}
\right\}
\eqno{(A15)}$$
We want to maximize $\lambda$ subject to the constraints given in equations
(A14) and (A15). It can be shown that ${\lambda}_{max} =
1/2$, and then $l = m = n = 0$. 
Thus the optimal disentanglement of the equatorial states can be achieved by
the linear map ${\tilde{T}}_{optimal}$, defined by ${\tilde{T}}_{optimal} (I) = I,
{\tilde{T}}_{optimal} ({\sigma}_1) = (1/2) {\sigma}_1, {\tilde{T}}_{optimal}
({\sigma}_3) = (1/2) {\sigma}_3, {\tilde{T}}_{optimal} ({\sigma}_2) = 0$, by
acting on the states of one of the parties. Obviously, ${\tilde{T}}_{optimal}$ is
a physical map, as it maps every density matrix of a two--level system to
another density matrix of the system \cite{explain}.   

Thus the map $\tilde{V}$ of equation (A8) (satisfying the condition
(A9) ) will optimally disentangle states of ${\cal
S}_{equator}$ iff $\Im \{\langle \overline{M_1}|M_0 \rangle\} = \Im \{\langle
M_0|M_1 \rangle\} = 0, \Re \{\langle \overline{M_1}|M_0 \rangle\} =
{\lambda}/(\lambda + 1)$ and $\lambda = 1/2$ (together with the other
conditions).


\begin{thebibliography}{99}
\bibitem{B1} Recently Bru{\ss} et. al. \cite{B2} pointed out the physical principles 
within quantum theory itself, which are responsible for the existence of optimal universal cloning 
machine.
\bibitem{greatdisc} By a ``great disc", we shall mean the disc of a great
circle.
\bibitem{whyitisexact} One may note here, that there is no set of states
${\rho}_{12}, {\rho}^{\prime}_{12}, \ldots$, where at least two of the reduced
density matrices $Tr_1[{\rho}_{12}], Tr_1[{\rho}^{\prime}_{12}], \ldots$ do not
commute with each other, but can be exactly disentangled by using a disentangling
machine on the states of party 2 \cite{Ghosh3}.
\bibitem{explain} In the derivation of the map ${\tilde{T}}_{optimal}$, we have
not used the condition of physicality of the map $\tilde{T}$, directly. But if
we use that condition explicitly, the set of all $(l, m, n)$, would contain the
set of all $(l, m, n)$ satisfying the conditions (A14) and
(A15). And so the maximum value of $\lambda$, that would be given
by the former set will be less than, in general, that given by the latter set
(which is $1/2$). So consideration of the latter set is appropriate for our
purpose.
\end{thebibliography}
\end{document}